\documentclass[sigconf,screen,nonacm]{acmart}
\AtBeginDocument{%
  }

\usepackage{float}
\usepackage{eso-pic}
\usepackage{tikz}
\usepackage{xcolor}
\usepackage{subcaption}

\renewcommand\footnotetextcopyrightpermission[1]{}

\newcommand{\rev}[1]{#1}
\newcommand{\method}{\rev{ContractSkill}}
\newcommand{\actionterm}[1]{#1}

\title{ContractSkill: Repairable Contract-Based Skills for Multimodal Web Agents}
\author{Zijian Lu}
\affiliation{%
  \institution{\rev{Nanjing University of Posts and Telecommunications}}
  \city{Nanjing}
  \country{China}}
\email{18818732360@163.com}

\author{Yiping Zuo}
\affiliation{%
  \institution{\rev{Nanjing University of Posts and Telecommunications}}
  \city{Nanjing}
  \country{China}}
\email{zuoyiping@njupt.edu.cn}

\author{Yupeng Nie}
\affiliation{%
  \institution{\rev{Nanjing University of Posts and Telecommunications}}
  \city{Nanjing}
  \country{China}}
\email{18094228550@163.com}

\author{Xin He}
\affiliation{%
  \institution{\rev{Nanjing University of Posts and Telecommunications}}
  \city{Nanjing}
  \country{China}}
\email{xhe@njupt.edu.cn}

\author{Weibei Fan}
\affiliation{%
  \institution{\rev{Nanjing University of Posts and Telecommunications}}
  \city{Nanjing}
  \country{China}}
\email{wbfan@njupt.edu.cn}

\author{\rev{Lianyong Qi}}
\affiliation{%
  \institution{\rev{China University of Petroleum (East China)}}
  \city{Qingdao}
  \country{China}}
\email{lianyongqi@gmail.com}

\author{\rev{Shi Jin}}
\affiliation{%
  \institution{\rev{Southeast University}}
  \city{Nanjing}
  \country{China}}
\email{jinshi@seu.edu.cn}

\renewcommand{\shortauthors}{}
\makeatletter
\def\@shortauthors{}
\def\@authorsaddresses{}
\makeatother
\authorsaddresses{}

\begin{abstract}
\rev{Self-generated skills for web agents are often unstable and can even hurt performance relative to direct acting. We argue that the key bottleneck is not only skill generation quality, but the fact that web skills remain implicit and therefore cannot be checked or locally repaired. To address this, we present } \method{}\rev{, a framework that converts a draft skill into an executable artifact with explicit procedural structure, enabling deterministic verification, fault localization, and minimal local repair. This turns skill refinement from full rewriting into localized editing of a single skill artifact. Experiments on VisualWebArena show that } \method{} \rev{is effective in realistic web environments, while MiniWoB provides a controlled test of the mechanism behind the gain. Under matched transfer layers, repaired artifacts also remain reusable after removing the source model from the loop, providing evidence of portability within the same benchmark family rather than full-benchmark generalization. These results suggest that the central challenge is not merely generating skills, but making them explicit, executable, and repairable. Code is available at \url{https://github.com/underfitting-lu/contractskill.git}.}
\end{abstract}



\begin{document}

\fancypagestyle{plain}{%
  \fancyhf{}
  \renewcommand{\headrulewidth}{0pt}
  \fancyfoot[C]{\thepage}}
\fancypagestyle{standardpagestyle}{%
  \fancyhf{}
  \renewcommand{\headrulewidth}{0pt}
  \fancyfoot[C]{\thepage}}
\pagestyle{plain}

\maketitle
\fancypagestyle{fancy}{%
  \fancyhf{}
  \renewcommand{\headrulewidth}{0pt}
  \fancyfoot[C]{\thepage}}
\pagestyle{fancy}
\markboth{}{}
\makeatletter
\gdef\shortauthors{}
\gdef\@shortauthors{}
\gdef\@authorsaddresses{}
\makeatother

\section{Introduction}
\rev{Long-horizon web tasks seem like exactly the setting where skills should help most. Yet recent evidence already points to a striking mismatch between that intuition and current practice. SkillsBench \cite{li2026skillsbench} reports that curated skills raise average pass rate by 16.2 percentage points, whereas self-generated skills provide no benefit on average . For us, this is the real motivating anomaly. It suggests that the problem is not whether external procedural knowledge can help agents, but whether model-written skills are stable enough to serve as executable objects in the first place.}

\rev{Our diagnosis is object-centered rather than pipeline-centered. The key bottleneck is not only skill generation quality, but the fact that web skills remain implicit and therefore cannot be checked or locally repaired. Most draft skills are loose textual objects: they omit explicit preconditions, blur step boundaries, leave success evidence underspecified, and provide no principled recovery policy when the page state diverges. Once execution fails, the agent typically rewrites the whole skill or abandons it altogether.}

\rev{This diagnosis also clarifies the gap with prior work. Reflection and self-repair methods update trajectories, memories, or transient plans in the current episode \cite{shinn2023reflexion,wang2024mobileagentv2}. Skill-bank style work maintains larger workflow collections or reusable knowledge repositories in open-ended settings \cite{wang2023voyager,chen2025pgagent}. Our target is different. We update the single skill artifact itself. The question we ask is whether a model-written web skill can be turned into an explicit object with preconditions, steps, postconditions, recovery rules, and termination checks, so that it becomes executable, diagnosable, locally repairable, and consumable by another model.}

\rev{We therefore propose } \method{}\rev{, which converts a draft skill into an executable contract and improves it through deterministic verification, fault localization, and minimal patch repair. Starting from a model-generated draft, } \method{} \rev{compiles the skill into an explicit artifact, executes it under a deterministic verifier, localizes failure to a specific step and error type, and edits only the implicated selector, condition, recovery rule, or action argument. Instead of treating failure as a reason to regenerate the whole skill, the framework treats it as a local repair problem over a persistent procedural object. The repaired artifact can then be reused by the source model or consumed by a target model without regenerating the skill.}

\rev{We test this claim from three complementary angles. First, on VisualWebArena, } \method{} \rev{improves execution in realistic multimodal web interaction where naive self-generated skills fail. Second, in MiniWoB, controlled comparisons show that the gain comes from deterministic verification, fault localization, and minimal repair rather than rewriting alone. Third, under matched transfer layers, repaired artifacts remain reusable across GLM-4.6V and Qwen3.5-Plus after removing the source model from the loop. This is a benchmark-specific test of portability rather than a claim of full-benchmark generalization.}

We summarize the primary contributions of this paper below.
\begin{itemize}
  \item \rev{We identify the failure of self-generated web skills as a representation problem rather than only a generation problem, motivated by prior evidence that self-generated skills often fail to deliver the gains achieved by curated procedural knowledge.}
  \item \rev{We propose } \method{}\rev{, an object-centered repair framework that converts a draft skill into an executable contract and improves the single skill artifact itself through deterministic verification, fault localization, and minimal patch operators.}
  \item \rev{We show that } \method{} \rev{is effective on VisualWebArena, while MiniWoB validates the mechanism under controlled conditions. We further show that matched transfer layers provide evidence that repaired artifacts remain reusable across models after removing the source model from the loop. This is a benchmark-specific portability claim rather than one of full-benchmark generalization.}
\end{itemize}

\begin{figure*}[t]
  \centering
  \includegraphics[width=0.98\textwidth]{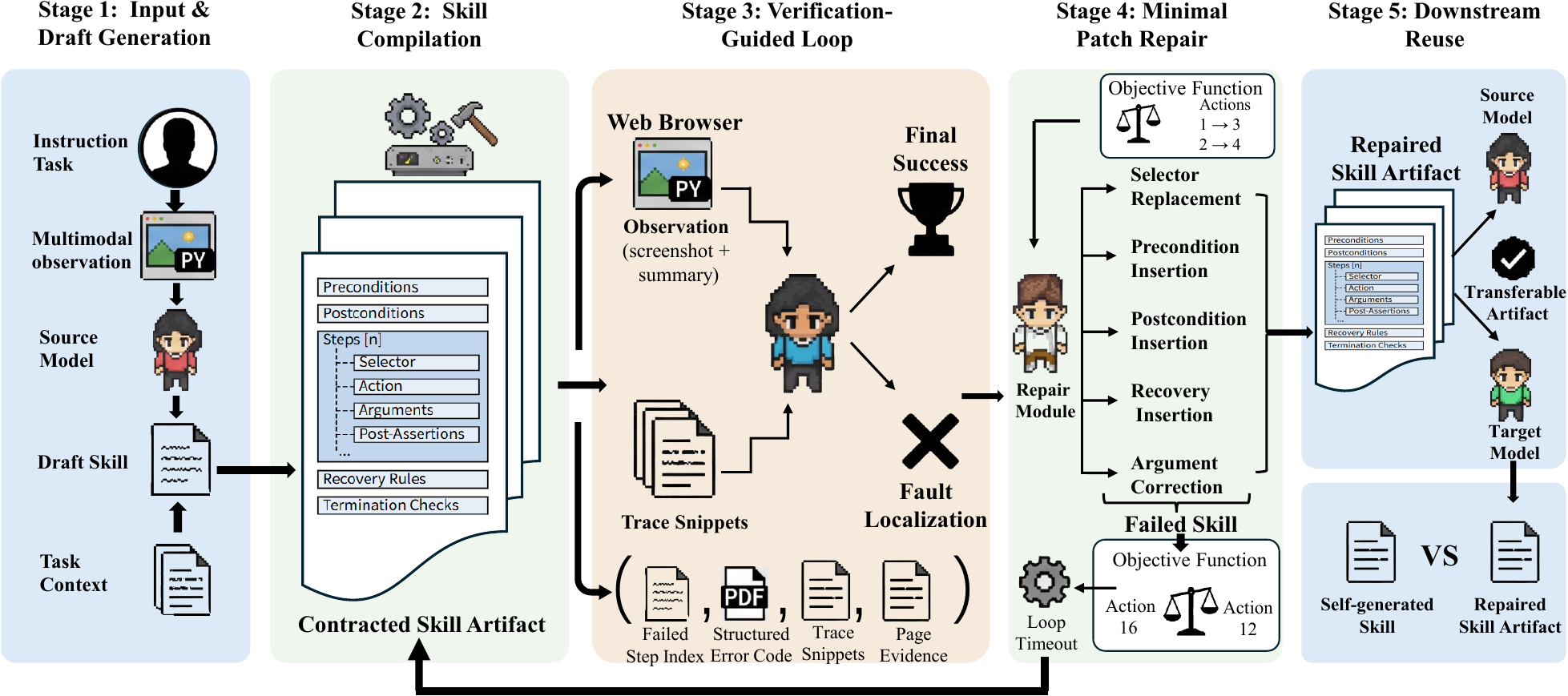}
  \caption{Overview of \method. The pipeline proceeds through five stages. It starts with input and draft generation, then compiles the draft into a contracted artifact, performs verification-guided execution with deterministic fault localization, applies minimal patch repair under an explicit objective, and finally reuses the repaired artifact with both the source model and a target model.}
  \Description{A five-stage pipeline for ContractSkill showing input and draft generation, skill compilation into a contracted artifact, verification-guided execution with deterministic fault localization, minimal patch repair, and downstream reuse by source and target models.}
  \label{fig:overview}
\end{figure*}

\rev{Figure~\ref{fig:overview} summarizes this artifact-centric workflow from draft generation to verifier-guided repair and downstream reuse.}

\section{Related Work}
\rev{We organize prior work around the reliability of self-generated web skills, focusing on why stronger skill generation alone does not guarantee executable and reusable skill artifacts, how prior work improves behavior online or externalizes procedure to narrow this gap, and what verifier-guided evidence is needed to make such artifacts auditable and repairable under a fixed execution protocol.}
\subsection{Multimodal Web Agents}
\rev{Recent benchmarks have rapidly expanded the scope of multimodal web agents. World of Bits and MiniWoB established controlled browser settings for basic interaction, while WebArena and VisualWebArena pushed evaluation toward longer-horizon, more realistic, and visually grounded web workflows \cite{shi2017worldofbits,liu2018reinforcement,zhou2023webarena,koh2024visualwebarena}. Other resources, including Mind2Web, WebLINX, BrowserGym, WorkArena++, SeeAct, AutoWebGLM, AgentOccam, and scaling-oriented agent training efforts, further broaden the empirical landscape \cite{deng2023mind2web,lu2024weblinx,chezelles2024browsergym,boisvert2025workarenapp,zheng2024seeact,lai2024autowebglm,yang2024agentoccam,trabucco2025internetscale}. These benchmarks are crucial for exposing realistic failure modes, but they mostly evaluate end-to-end agent behavior rather than how a self-generated web skill should be represented, verified, and repaired after execution breaks.}

\subsection{Self-Repair and Reflection in Agents}
\rev{Within these environments, self-repair and reflection methods improve agent behavior by revising trajectories, reasoning traces, memories, or policies after observing failure \cite{shinn2023reflexion,wang2024mobileagentv2,yao2026parammem,weng2026gea,yao2023react,zhao2023expel}. These literatures are highly relevant for web agents because they shows that feedback-driven refinement can materially improve performance in interactive environments. However, the object being updated is usually the agent's ongoing behavior or internal state in the current episode. By contrast, we repair an externalized skill artifact itself so that the result persists as a reusable object across episodes and can also be consumed by another model.}

\subsection{Skill Artifacts and Procedural Knowledge}
\rev{A complementary line of work treats skills as explicit, reusable procedural knowledge rather than transient prompting context \cite{xu2026agentskills,li2026skillsbench,li2026singleagentskills}. Agent Workflow Memory stores reusable textual workflows, ASI induces and verifies executable programmatic skills online, PolySkill separates abstract goals from concrete implementations for cross-site reuse, SkillWeaver discovers website-specific APIs, Hierarchical Memory Tree organizes web experience with explicit pre/post conditions, and RAG-GUI retrieves external GUI guidance at inference time \cite{wang2024awm,wang2025asi,yu2025polyskill,zheng2025skillweaver,tan2026hmt,xu2025raggui}. This line is the closest to ours, but most prior work studies reusable knowledge at the level of workflow collections, APIs, or broader memory structures. Our focus is narrower and more operational: a single model-written multimodal web skill that must become executable, diagnosable, minimally repairable, and reusable across models.}

\subsection{Program Repair and Verifier-Guided Optimization}
\rev{Once the skill itself becomes the repair target, program repair provides the methodological lens for our framework. Classic work such as GenProg, PAR, SemFix, DirectFix, Angelix, and Prophet emphasizes three ideas that are equally central here: localize the failure, constrain the edit space, and use a verifier to decide whether a patch should be accepted \cite{legoues2012genprog,kim2013par,nguyen2013semfix,mechtaev2015directfix,mechtaev2016angelix,long2016prophet}. Verifier-guided refinement frameworks such as counterexample-guided abstraction refinement further show how failed checks can drive iterative improvement \cite{clarke2000cegar}. We borrow these principles not to repair source code directly, but to repair executable web-skill artifacts whose correctness is grounded in observable page-state evidence.}

\section{Problem Definition}
\subsection{Environment}
We study a multimodal web environment similar to VisualWebArena, WorkArena, and OSWorld \cite{koh2024visualwebarena,drouin2024workarena,xie2024osworld}. An agent receives a natural-language task instruction $x$. At step $t$, the environment provides an observation $o_t = (I_t, D_t)$, where $I_t$ is the current page screenshot and $D_t$ is a structured page summary, such as \rev{a Document Object Model (DOM) abstraction}, accessibility summary, or textual page state description. The agent outputs an action $a_t \in \mathcal{A}$, where $\mathcal{A}$ includes operations such as \actionterm{click}, \actionterm{type}, \actionterm{scroll}, and \actionterm{select}. The environment then transitions to a new observation $o_{t+1}$.

Task success is judged by a deterministic verifier $V$. Unlike free-form language judgments, $V$ checks final and intermediate page states using reproducible signals, such as URL patterns, DOM matches, text presence, form values, or page-state transitions. This design is consistent with evaluation practice in reproducible web-agent benchmarks, which ground success in observable page evidence rather than subjective model preferences \cite{zhou2023webarena,koh2024visualwebarena,drouin2024workarena}. This setting lets us reason about failure in programmatic terms rather than relying only on subjective model reflection.

\subsection{Draft Skill}
Given instruction $x$ and an initial observation, a source model first generates a textual draft skill $s_0$. This draft is a weak procedural object because it contains useful action intent but is not yet executable in a controlled or verifiable way. In particular, $s_0$ may omit hidden state assumptions, under-specify selectors, conflate goals with actions, or fail to state what evidence marks step completion. This pattern is common in self-generated manuals, mobile skills, and benchmarked skill artifacts \cite{zhang2023appagent,chen2024automanual,li2026skillsbench}. We therefore treat the draft as a starting point rather than as the final reusable skill.

\subsection{Contracted Skill Artifact}
The core object in this paper is the contracted skill artifact $s$. We formalize it as
\begin{equation}
s=\bigl(g, P, U, Q, R, T\bigr), \qquad
U=\{u_j\}_{j=1}^{m},
\end{equation}
where $g$ is the goal, $P$ denotes preconditions, $U$ is the ordered step list, $Q$ denotes postconditions, $R$ denotes recovery rules, and $T$ denotes termination checks. Each step is itself a structured object,
\begin{equation}
u_j=\bigl(\mathrm{sel}_j,\mathrm{act}_j,\mathrm{arg}_j,\Pi_j\bigr),
\end{equation}
where $\mathrm{sel}_j$ is the selector, $\mathrm{act}_j$ is the action type, $\mathrm{arg}_j$ denotes optional arguments, and $\Pi_j$ is the set of post-assertions for step $j$. This representation makes the skill executable at the step level and gives the verifier explicit hooks for diagnosis. It also makes the skill closer in spirit to reusable external knowledge objects, such as interaction manuals and structured prior knowledge stores, than to a one-off prompt fragment \cite{chen2024automanual,wang2023voyager,chen2025pgagent}.

\subsection{Objective}
Starting from draft skill $s_0$, we search over a local neighborhood of candidate repairs and seek an improved artifact $s^\ast$ that balances success, cost, and edit size according to the following objective.
\begin{equation}
s^\ast = \arg\max_{s \in \mathcal{N}_K(s_0)}
\mathrm{Succ}(s) - \lambda \mathrm{Cost}(s) - \gamma \mathrm{Edit}(s,s_0).
\end{equation}
Here $\mathrm{Succ}(s)$ denotes task success under the verifier, $\mathrm{Cost}(s)$ captures execution cost such as action count or tool calls, and $\mathrm{Edit}(s,s_0)$ penalizes unnecessarily large modifications. This objective reflects the desired behavior of a repair system, which is to improve the skill while preserving as much of the original procedural content as possible. The edit penalty is deliberately aligned with minimal-change preferences that have been useful in program repair.

\section{ContractSkill}
\subsection{Skill Artifact Schema}
\method\ represents each skill as an explicit artifact with the fields skill\_name, goal, preconditions, steps, postconditions, recovery, and terminate. Each step contains four mandatory elements: selector, action, args, and post\_assertion. Compared with free-form text, this schema exposes the exact operational assumptions behind the skill. The schema is designed to externalize procedures as reusable objects, which is also the motivating direction behind manual construction and page-level knowledge extraction in recent agent systems.

\subsection{Deterministic Verifier}
We distinguish two verification levels. A \rev{final verifier} determines whether the full task has succeeded, and a \rev{step verifier} checks whether each action was executed under valid state assumptions and produced the expected page evidence. Let $\mathcal{T}$ denote the execution trace together with the page-state trace. We write the verifier as
\begin{equation}
\mathcal{V}(s,\mathcal{T})=
\begin{cases}
1, & \text{if the task succeeds,}\\
(0,i,e,\tau_i), & \text{otherwise,}
\end{cases}
\end{equation}
where $i$ is the failed step index, $e$ is a structured error code, and $\tau_i$ is the local trace evidence used for diagnosis. Verification uses deterministic checks such as URL matching, DOM retrieval, textual assertions, form-value checks, and state-change predicates. In practice, selectors are resolved on the current accessibility or DOM snapshot using role, visible text, and stable attributes rather than persistent node ids. Postconditions are instantiated from action-specific templates, such as value matching for \actionterm{type}, state or navigation change for \actionterm{click}, and option validation for \actionterm{select}. This design keeps the repair signal grounded in observable page state rather than in model self-evaluation, matching the execution-oriented evaluation logic used in WebArena, VisualWebArena, and WorkArena \cite{zhou2023webarena,koh2024visualwebarena,drouin2024workarena}. \rev{Table~\ref{tab:error-codes} summarizes the structured error codes returned by the verifier.}

\begin{table}[t]
  \caption{Structured error codes used by the deterministic verifier.}
  \label{tab:error-codes}
  \centering
  \small
  \begin{tabular}{lp{0.55\columnwidth}}
    \toprule
    Error code & Meaning \\
    \midrule
    \texttt{NOT\_FOUND} & The specified selector does not match a valid page element. \\
    \texttt{WRONG\_STATE} & The current page state violates a step precondition. \\
    \texttt{ASSERT\_FAIL} & The action completes, but the expected postcondition is not satisfied. \\
    \texttt{LOOP\_TIMEOUT} & The agent repeats unproductive actions or exceeds the step budget. \\
    \texttt{INPUT\_INVALID} & The action arguments do not satisfy page-side constraints. \\
    \bottomrule
  \end{tabular}
\end{table}

\subsection{Fault Localization}
When execution fails, \method\ does not treat the failure as a monolithic property of the entire skill. Instead, it extracts a localized diagnosis
\begin{equation}
(i,e,\tau_i)=\mathrm{Localize}\bigl(s,\mathcal{T}\bigr),
\end{equation}
where $i$ is the failed step index, $e$ is the structured error code, and $\tau_i$ is the relevant trace snippet or page-state evidence. This step-level localization is important because different failures imply different repair actions. A missing selector and an invalid state assumption should not trigger the same patch strategy. The same diagnosis-first discipline is central in semantics-based repair methods, which isolate a small repair site before synthesis or validation \cite{nguyen2013semfix,mechtaev2016angelix}.

\subsection{Minimal Patch Operators}
We define five local patch operators,
\begin{equation}
\mathcal{P}=
\begin{aligned}
\{&\textsc{SelReplace},\textsc{PreInsert},\textsc{PostInsert},\\
  &\textsc{RecoveryInsert},\textsc{ArgCorrect}\}.
\end{aligned}
\end{equation}
Using this operator set, the $K$-step repair neighborhood of the draft artifact is
\begin{equation}
\mathcal{N}_K(s_0)=
\left\{
p_k\circ \cdots \circ p_1(s_0)\; \middle|\;
p_r\in\mathcal{P},\; 0\le k\le K
\right\}.
\end{equation}
Each operator changes only the artifact fragment tied to the diagnosed failure. This design preserves the rest of the artifact and avoids the instability that often accompanies full-text rewrites. The operator set intentionally favors small, interpretable edits, echoing the minimality bias emphasized in DirectFix and learning-guided repair systems such as Prophet \cite{mechtaev2015directfix,long2016prophet}. \rev{Table~\ref{tab:patch-operators} summarizes these minimal patch operators.}

\begin{table}[t]
  \caption{Minimal patch operators in \method.}
  \label{tab:patch-operators}
  \centering
  \small
  \begin{tabular}{lp{0.53\columnwidth}}
    \toprule
    Operator & Typical use \\
    \midrule
    Selector replacement & Replace or expand a brittle selector after \texttt{NOT\_FOUND}. \\
    Precondition insertion & Add a missing state assumption after \texttt{WRONG\_STATE}. \\
    Postcondition insertion & Add missing evidence of completion after \texttt{ASSERT\_FAIL}. \\
    Recovery insertion & Add a fallback action such as close-popup or reload. \\
    Argument correction & Fix input values or action arguments after \texttt{INPUT\_INVALID}. \\
    \bottomrule
  \end{tabular}
\end{table}

\subsection{Verifier-Guided Repair Loop}
Algorithmically, \method\ follows an iterative execute-diagnose-patch-validate loop. First, the source model produces a draft skill and compiles it into an artifact. Second, the artifact is executed on a repair set. Third, the verifier reports the failing step and error code. Fourth, the repair module chooses one or more minimal patch operators and constructs candidate patches. Finally, each candidate is validated under the same verifier and accepted only if it improves the current score. Concretely, we write
\begin{equation}
J(s)=\mathrm{Succ}(s)-\lambda \mathrm{Cost}(s)-\gamma \mathrm{Edit}(s,s_0),
\end{equation}
\begin{equation}
\mathrm{Accept}(s',s)=
\begin{cases}
1, & \text{if } J(s') > J(s),\\
0, & \text{otherwise.}
\end{cases}
\end{equation}
In our setup, each round is limited to a small edit budget, typically one to three patches, so that repair remains local and interpretable. The acceptance logic follows the broader verifier-guided refinement pattern in which counterexamples or failed checks narrow the next search step \cite{clarke2000cegar}.

\begin{figure*}[t]
  \centering
  \includegraphics[width=0.98\textwidth]{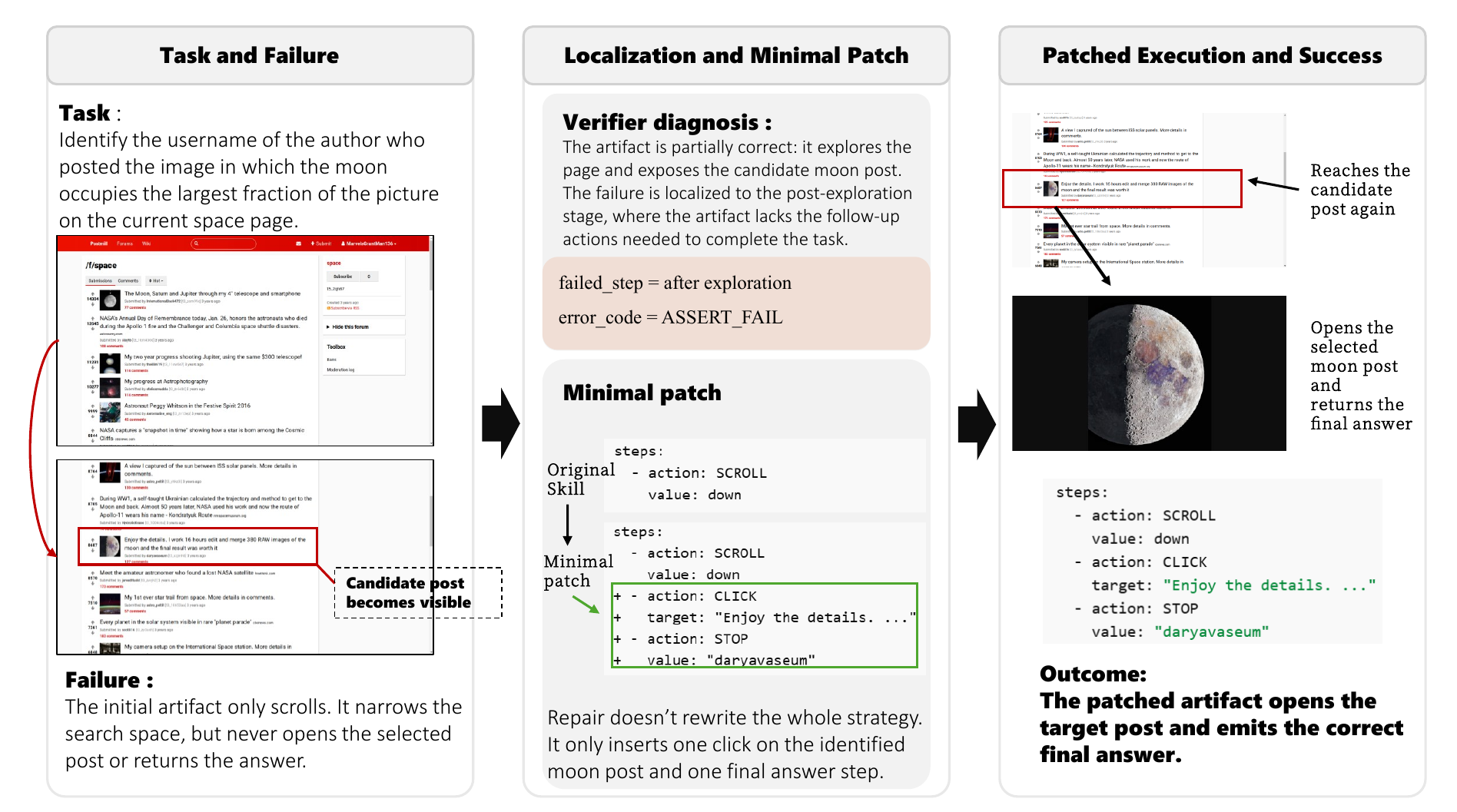}
  \caption{Representative repair case on VisualWebArena. The verifier localizes the failure to the post-exploration stage, and \method\ repairs the artifact by inserting only the missing continuation steps rather than rewriting the whole strategy.}
  \Description{A qualitative case study showing task failure, verifier localization, minimal patch insertion, and successful patched execution on a VisualWebArena task.}
  \label{fig:repair-case}
\end{figure*}

\section{Experimental Setup}
Our experiments are organized around three questions. We ask (1) whether \method\ improves end-to-end web execution over direct acting and naive skill use, (2) whether any improvement comes specifically from verifier-guided local repair rather than rewriting alone, and (3) whether repaired skill artifacts remain reusable by another model under matched transfer layers after the source model is removed from the loop. \rev{Across all experiments, we treat efficiency as a success-cost trade-off rather than strict dominance over every baseline: the point is to test whether modest extra execution cost can buy substantially higher success when brittle self-generated skills fail.} We structure the evaluation accordingly.

\subsection{Benchmarks and Evaluation Splits}
We evaluate on two benchmarks with complementary roles. VisualWebArena (VWA) \cite{koh2024visualwebarena} serves as the realistic multimodal web benchmark, where long-horizon interaction and environment noise make brittle skills particularly costly. We report VWA results on the cleaned \rev{main\_100} subset, i.e., the 100 tasks for which matched summaries are available across all compared runs. For MiniWoB \cite{liu2018reinforcement}, we use the \rev{main\_20} handbook split, which contains 20 tasks with 10 instances per task. We evaluate both benchmarks over three runs. MiniWoB serves as a controlled benchmark with simpler task structure, allowing the effect of repair decisions to be isolated more cleanly. This yields 300 task-level evaluations on VWA and 600 evaluation episodes on MiniWoB in total.

For transfer, we replace the usual seen-versus-held-out template split with benchmark-specific transfer layers that better match our claim. \rev{Layer 1} (layer1\_paired\_success) contains tasks that both source models originally solve, enabling paired comparisons under matched difficulty. \rev{Layer 2} (layer2\_source\_only\_success) contains tasks solved by only one source model, which directly tests whether a repaired skill can carry useful procedural knowledge beyond the overlap of easy cases. We report both directional settings, Q$\rightarrow$G and G$\rightarrow$Q.

\subsection{Models}
We instantiate the source and target agents with GLM-4.6V and Qwen3.5-Plus. In the self-use setting, the same model generates the draft skill, repairs it, and executes the repaired artifact. In the transfer setting, the source model produces and repairs the artifact, while the target model consumes it without regenerating the skill. This separation lets us test whether the final artifact encodes reusable procedure rather than model-specific hidden context.
\subsection{Baselines}
We compare against two core baselines in all main-result tables. \rev{No-Skill} executes the task directly without any external procedural guidance, representing the standard direct-acting setting. \rev{Self-Generated Skill} lets the acting model produce a skill and immediately use it, but does not perform verifier-guided repair after failure. It therefore isolates the effect of skill writing alone, without the additional benefit of explicit verification or repair. On MiniWoB, we additionally include more controlled comparisons. \rev{Text-Only Rewrite} rewrites the skill after failure in free-form text, without contracts, structured error codes, or localized edits, and therefore isolates the effect of rewriting alone. We also report two ablations of \method. \rev{No Failure Localization} removes targeted identification of the failing skill component, and \rev{Unconstrained Repair} permits broad revisions rather than restricted local fixes. These controlled variants help determine whether the gains of \method\ arise from structured verifier-guided repair.

\subsection{Metrics}
Our primary metric is verifier-based success rate. \rev{Under this framing,} we also report average steps and average total tokens whenever those fields are present in the cleaned summaries. A repaired artifact may spend extra steps or tokens to complete tasks that a brittle self-generated skill fails to finish. Tokens are reported as average total tokens per task and abbreviated in thousands (k). For transfer, we compare repaired-artifact execution against the target model's own self-generated skill baseline and report both absolute success and the point improvement.

\subsection{Implementation Details}
All compared methods use the same observation interface, action space, and deterministic verifier. For fairness, we keep the execution budget and the maximum allowed repair effort fixed across methods within each benchmark family. The verifier evaluates structured environment state through deterministic checks, including URL consistency, visible-text matching, form-value validation, and retrieval over the DOM summary. We re-resolve selectors on the latest page snapshot after every action and allow a short wait-and-requery window for delayed rendering, which reduces brittleness to dynamic interfaces without introducing non-deterministic matching. This controlled setup makes the comparison method-centric. Gains cannot be attributed to extra interaction budget, different action primitives, or weaker success criteria.

\section{Main Results}
We report the main results in the order of argumentative strength. We first test whether \method\ works in realistic web interaction, then examine the same claim in a controlled environment where the mechanism is easier to isolate, and finally test whether the repaired artifact remains reusable under matched transfer layers after the source model is removed from the loop.

\subsection{VisualWebArena}
Table~\ref{tab:vwa-main} reports the main VWA results. The overall pattern is clear. On realistic web tasks, simply having a skill is not enough, but repairing it into a contracted artifact is consistently beneficial. With GLM-4.6V, ContractSkill reaches \rev{28.1\%} success, compared with \rev{12.5\%} for No-Skill and \rev{9.4\%} for Self-Generated Skill. The efficiency profile also improves. Average steps drop from 12.59 for No-Skill to 3.62, and average token usage decreases from 52.7k to 46.45k. With Qwen3.5-Plus, the packaged final verified aggregation raises ContractSkill to \rev{37.5\%}, exceeding both No-Skill (\rev{28.1\%}) and Self-Generated Skill (\rev{10.9\%}), while keeping average token usage slightly below No-Skill (40.1k vs.\ 42.6k). On VWA, self-generated skills can underperform direct acting because an unverified but committed procedure often fails early and suppresses corrective exploration. As a result, brittle skills may use fewer steps and tokens precisely because they commit to the wrong procedure too soon. These results matter because VWA is precisely the regime in which brittle skills tend to fail, with long trajectories, changing page state, and multimodal ambiguity. The fact that ContractSkill improves both models suggests that the gain comes from structured repair of the external artifact rather than from model-specific prompting tricks.

\begin{table}[t]
  \caption{Main results on VisualWebArena main\_100.}
  \label{tab:vwa-main}
  \centering
  \small
  \renewcommand{\arraystretch}{1.08}
  \begin{tabular*}{\columnwidth}{@{\extracolsep{\fill}}lccc}
    \toprule
    \multicolumn{4}{c}{GLM-VWA} \\
    \midrule
    Method & Succ. (\%) & Steps & Tokens (k) \\
    \midrule
    No-Skill & 12.5 & 12.59 & 52.7 \\
    Self-Generated Skill & 9.4 & 1.95 & 5.6 \\
    ContractSkill & \textbf{28.1} & 3.62 & 46.45 \\
    \midrule
    \multicolumn{4}{c}{Qwen-VWA} \\
    \midrule
    Method & Succ. (\%) & Steps & Tokens (k) \\
    \midrule
    No-Skill & 28.1 & 9.41 & 42.6 \\
    Self-Generated Skill & 10.9 & 1.91 & 7.3 \\
    ContractSkill & \textbf{37.5} & 8.52 & 40.1 \\
    \bottomrule
  \end{tabular*}

  \vspace{2pt}
  \parbox{\columnwidth}{\footnotesize \textit{Note.} Tokens are average total tokens per task reported in thousands.}
\end{table}

\subsection{MiniWoB}
Table~\ref{tab:miniwob-main} provides the controlled counterpart to the VWA findings. Here the conclusion is even cleaner. ContractSkill is the strongest method for both models, reaching \rev{77.5\%} on GLM and \rev{81.0\%} on Qwen. Relative to Self-Generated Skill, the improvement is 11.0 points for GLM and 20.5 points for Qwen. The efficiency numbers show that this gain is not achieved by reverting to expensive direct exploration. On GLM, ContractSkill uses 1.93 average steps and 5.16k tokens, remaining far below the 4.11-step cost of No-Skill. On Qwen, ContractSkill increases steps only slightly over Self-Generated Skill (2.00 vs.\ 1.85) while producing a much larger success gain. Because MiniWoB removes much of the uncontrolled noise present in open web environments, these results strengthen the central claim that the improvement comes from repairable external procedure, not merely from extra interaction budget. We place Text-Only Rewrite in the ablation table below, where it serves as the most direct check against the ``it is just rewriting'' alternative explanation.

\begin{table}[t]
  \caption{Main results on MiniWoB main\_20 with 10 instances per task.}
  \label{tab:miniwob-main}
  \centering
  \small
  \renewcommand{\arraystretch}{1.08}
  \begin{tabular*}{\columnwidth}{@{\extracolsep{\fill}}lccc}
    \toprule
    \multicolumn{4}{c}{GLM-MiniWoB} \\
    \midrule
    Method & Succ. (\%) & Steps & Tokens (k) \\
    \midrule
    No-Skill & 61.5 & 4.11 & 4.44 \\
    Self-Generated Skill & 66.5 & 1.84 & 4.6 \\
    ContractSkill & \textbf{77.5} & 1.93 & 5.16 \\
    \midrule
    \multicolumn{4}{c}{Qwen-MiniWoB} \\
    \midrule
    Method & Succ. (\%) & Steps & Tokens (k) \\
    \midrule
    No-Skill & 56.5 & 1.61 & 1.9 \\
    Self-Generated Skill & 60.5 & 1.85 & 4.5 \\
    ContractSkill & \textbf{81.0} & 2.00 & 7.9 \\
    \bottomrule
  \end{tabular*}

  \vspace{2pt}
  \parbox{\columnwidth}{\footnotesize \textit{Note.} Tokens are average total tokens per task reported in thousands.}
\end{table}

\subsection{\rev{Portability Under Matched Transfer Layers}}
\rev{Table~\ref{tab:transfer} tests a narrower claim than broad cross-benchmark transfer: whether the repaired artifact remains useful after the source model is removed from the loop under benchmark-specific matched transfer layers. The results provide positive evidence for that claim.} Aggregated over both directions, transfer raises VWA success from \rev{32.6\%} to \rev{80.4\%} (+47.8 points) and MiniWoB success from \rev{83.5\%} to \rev{96.2\%} (+12.8 points) relative to the target model's own self-generated skill baseline. The directional results show the same pattern rather than a one-sided effect. On VWA, Q$\rightarrow$G improves from 29.2\% to 79.2\%, and G$\rightarrow$Q improves from 36.4\% to 81.8\%. On MiniWoB, Q$\rightarrow$G reaches 100.0\% versus 87.5\%, and G$\rightarrow$Q reaches 92.3\% versus 79.2\%. Transfer execution remains lightweight, requiring only 1.77--3.21 average steps depending on the direction, and the released summaries record zero live token usage because the artifact is reused instead of regenerated online. \rev{The main takeaway is not unrestricted transfer, but artifact reuse after removing the source model from the loop. The Layer 2 gains are especially informative because they show reuse beyond the easy overlap region within this matched setting, even though the evaluation still stays within the same benchmark family.}

\begin{table*}[t]
  \caption{\rev{Portability under matched transfer layers} compared against the target model's own self-generated skill baseline. Aggregate rows merge Layer 1 and Layer 2 within each benchmark.}
  \label{tab:transfer}
  \centering
  \small
  \renewcommand{\arraystretch}{1.08}
  \begin{tabular}{lccccc}
    \toprule
    Setting & Transfer Succ. (\%) & Target Skill (\%) & Delta & Transfer Steps & Baseline Steps \\
    \midrule
    VWA (aggregate) & \textbf{80.4} & 32.6 & +47.8 & 2.83 & 1.46 \\
    VWA Q$\rightarrow$G & 79.2 & 29.2 & +50.0 & 3.21 & 1.46 \\
    VWA G$\rightarrow$Q & 81.8 & 36.4 & +45.5 & 2.41 & 1.45 \\
    MiniWoB (aggregate) & \textbf{96.2} & 83.5 & +12.8 & 1.80 & 1.66 \\
    MiniWoB Q$\rightarrow$G & 100.0 & 87.5 & +12.5 & 1.84 & 1.68 \\
    MiniWoB G$\rightarrow$Q & 92.3 & 79.2 & +13.1 & 1.77 & 1.63 \\
    \bottomrule
  \end{tabular}
\end{table*}

\section{Analysis and Ablation}
We next ask why the main gains appear. Since the MiniWoB bundles provide the most complete controlled ablations, we use them to isolate the contribution of structured repair.

\subsection{Failure Localization and Minimal Repair}
Table~\ref{tab:ablation} supports two conclusions. First, the gain is not explained by rewriting alone. Text-Only Rewrite reaches only 62.0\% on GLM and 60.5\% on Qwen, far below full ContractSkill at 77.5\% and 81.0\%, respectively. This gap is important because Text-Only Rewrite already gives the model an opportunity to revise the skill after observing failure; what it lacks is the contracted artifact, explicit verifier feedback, and localized repair target.

Figure~\ref{fig:repair-case} shows the same mechanism qualitatively on a representative VWA task. In this example, the original artifact already narrows the search space and surfaces the candidate moon post, but it stops at the exploration stage and never performs the final task-completing actions. The verifier localizes this failure as a missing continuation after exploration, and the repair module inserts only the missing click and answer steps. The repaired artifact therefore succeeds without changing the rest of the strategy.

Second, both design choices in \method\ matter. Removing failure localization drops performance to 65.0\% on GLM and 70.0\% on Qwen, indicating that knowing \rev{where} the artifact fails is a major part of the benefit. Relaxing the repair objective into unconstrained editing performs slightly better than removing localization, but still reaches only 68.5\% on GLM and 70.5\% on Qwen. The efficiency trend points in the same direction. On Qwen, unconstrained repair increases average token usage from 7.9k to 11.3k without recovering the lost success. On GLM, it also increases token usage beyond both Text-Only Rewrite and the localization ablation. Taken together, these results support the intended mechanism of \method. Failure localization narrows the repair site, and minimal patching prevents costly global rewrites that are harder to validate and easier to destabilize.

\begin{table*}[t]
  \caption{MiniWoB ablation results for fault localization and repair constraints.}
  \label{tab:ablation}
  \centering
  \small
  \begin{tabular}{lcccccc}
    \toprule
    & \multicolumn{3}{c}{GLM-MiniWoB} & \multicolumn{3}{c}{Qwen-MiniWoB} \\
    \cmidrule(lr){2-4}\cmidrule(lr){5-7}
    Variant & Success (\%) & Avg.\ Steps & Avg.\ Tokens (k) & Success (\%) & Avg.\ Steps & Avg.\ Tokens (k) \\
    \midrule
    Full ContractSkill & \textbf{77.5} & 1.93 & 5.16 & \textbf{81.0} & 2.00 & 7.9 \\
    Unconstrained Repair & 68.5 & 1.92 & 6.6 & 70.5 & 2.00 & 11.3 \\
    No Failure Localization & 65.0 & 1.75 & 4.9 & 70.0 & 1.95 & 7.4 \\
    Text-Only Rewrite & 62.0 & 1.71 & 2.7 & 60.5 & 1.70 & 5.3 \\
    \bottomrule
  \end{tabular}

  \vspace{2pt}
  \parbox{0.98\textwidth}{\footnotesize \textit{Note.} Tokens are average total tokens per task reported in thousands.}
\end{table*}

\subsection{Template-Level MiniWoB Patterns}
\begin{figure}[t]
  \centering
  \includegraphics[width=0.82\columnwidth,trim=6bp 13bp 0bp 16bp,clip]{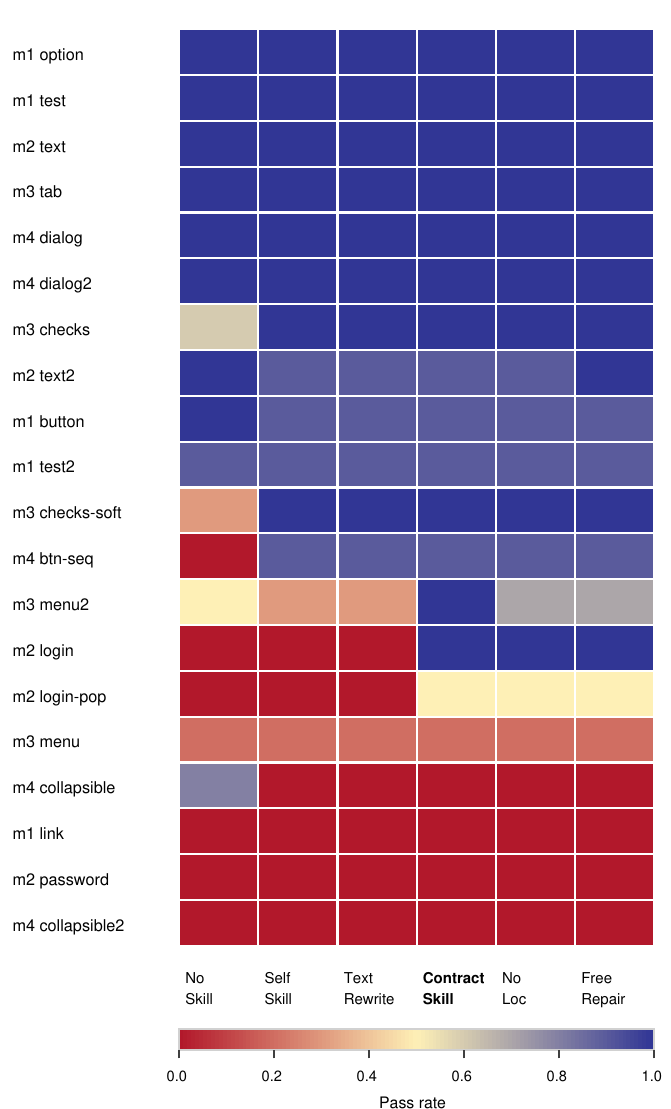}
  \caption{Template-level MiniWoB pass rates across the main baselines and ablations. ContractSkill gains concentrate on a small subset of repairable templates.}
  \Description{A heatmap of MiniWoB template-level pass rates for no-skill, skill-no-repair, text-only rewrite, ContractSkill, ContractSkill without failure localization, and ContractSkill with unconstrained repair. Most easy templates are saturated, while the largest gains of ContractSkill appear on a smaller set of harder templates.}
  \label{fig:miniwob-heatmap}
\end{figure}

Figure~\ref{fig:miniwob-heatmap} adds template-level pass rates to the aggregate MiniWoB ablation. Many easy templates are already saturated across all methods, so the overall gain does not come from uniform improvement everywhere. Instead, the advantage of full ContractSkill concentrates on a smaller set of templates that require reliable continuation after partial progress. The clearest gains appear on the two login templates and on mw\_\allowbreak m3\_\allowbreak click\_\allowbreak menu\_\allowbreak 2, where the repaired artifact clearly outperforms both skill\_\allowbreak no\_\allowbreak repair and text\_\allowbreak only\_\allowbreak rewrite.

The same figure also clarifies the remaining boundary cases. Some templates remain unsolved for every method, including mw\_\allowbreak m1\_\allowbreak click\_\allowbreak link and mw\_\allowbreak m2\_\allowbreak enter\_\allowbreak password. A different failure mode appears on mw\_\allowbreak m4\_\allowbreak click\_\allowbreak collapsible, where no-skill execution outperforms all skill-based variants. This pattern is consistent with the mechanism suggested by Table~\ref{tab:ablation}. Structured repair helps most when there is a reusable procedural skeleton to preserve and patch, but it is not a universal substitute for exploration on every template.

\begin{figure}[t]
  \centering
  \begin{subfigure}{\columnwidth}
    \centering
    \makebox[\linewidth][c]{\includegraphics[width=1.04\linewidth]{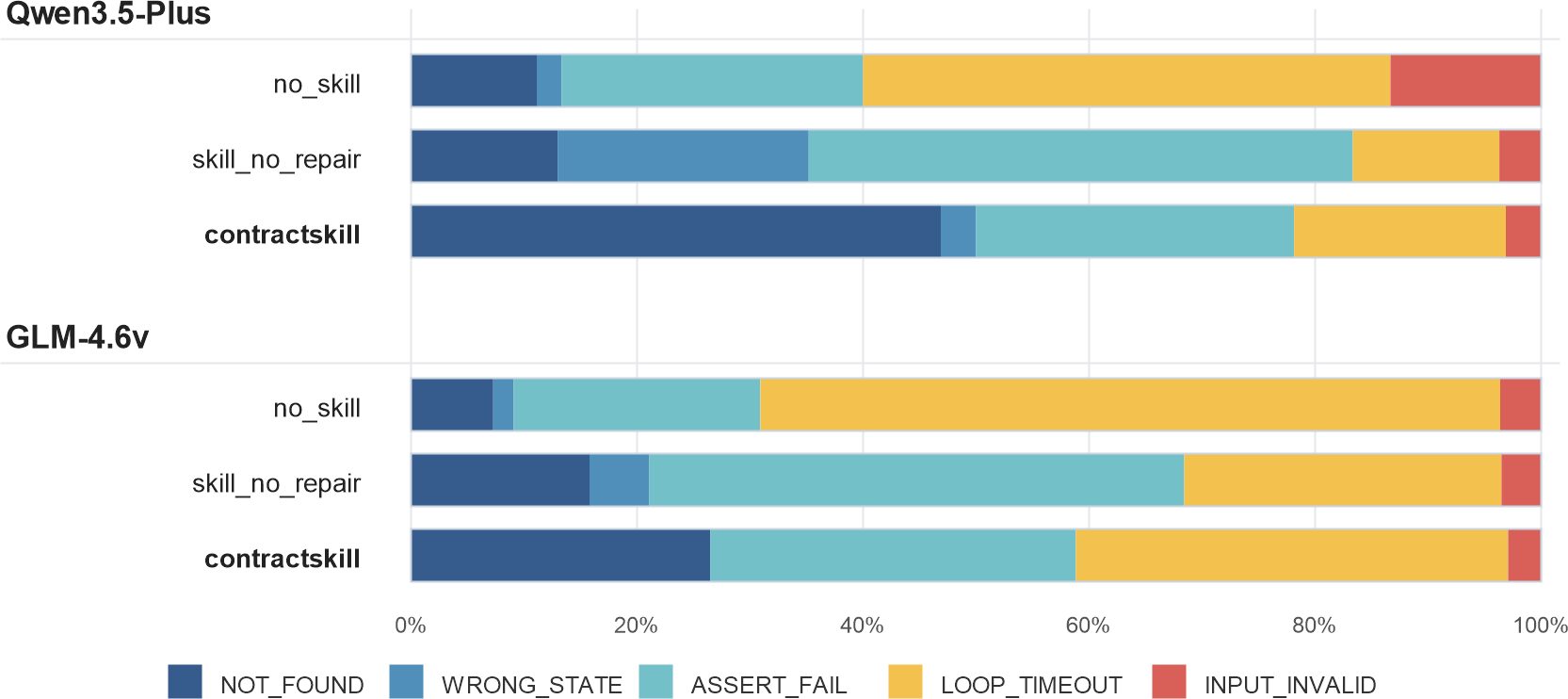}}
    \caption{\rev{Verifier error codes among failed episodes.}}
  \end{subfigure}

  \vspace{4pt}
  \begin{subfigure}{\columnwidth}
    \centering
    \makebox[\linewidth][c]{\includegraphics[width=1.04\linewidth]{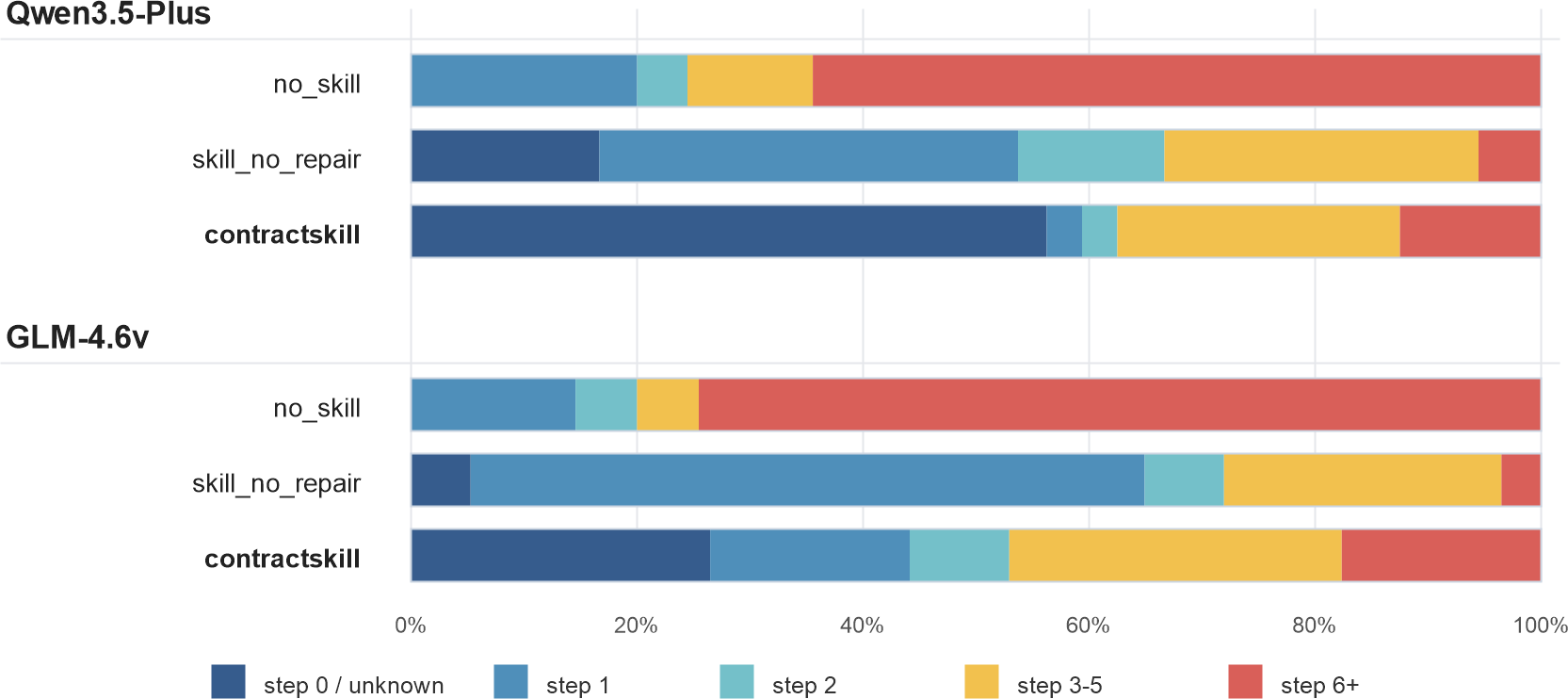}}
    \caption{\rev{First failing step bucket among failed episodes.}}
  \end{subfigure}
  \caption{\rev{VWA failure analysis across Qwen3.5-Plus and GLM-4.6V. Across both models, ContractSkill reduces the dominance of late coarse failures that are common in No-Skill execution and exposes a larger share of residual failures as earlier, more localizable breakdowns.}}
  \Description{Two stacked horizontal bar charts for VisualWebArena arranged vertically. The top chart shows failure-code distributions for no-skill, self-generated skill, and ContractSkill under Qwen3.5-Plus and GLM-4.6V. The bottom chart shows the corresponding first-failing-step distributions. ContractSkill shifts failures away from late coarse outcomes and toward more localized failure categories.}
  \label{fig:vwa-failure-supp}
\end{figure}

\rev{Figure~\ref{fig:vwa-failure-supp} summarizes the residual failure structure across both source models in VisualWebArena. Panel~(a) shows verifier error codes and Panel~(b) shows the first failing step bucket. Relative to No-Skill, ContractSkill leaves fewer failures concentrated in broad late-stage profiles, especially step-6+ outcomes and loop-timeout-heavy mixes, and exposes more failures as localized selector or assertion errors. Manual inspection suggests that self-generated skills commonly fail through early wrong selectors, missing preconditions, and premature stop conditions. These committed but brittle procedures often truncate exploration before the agent can recover, which helps explain why Self-Generated Skill can trail direct acting while still using fewer steps and tokens. This shift does not remove all failures, but it makes the remaining ones more structured and more informative for targeted repair.}

\section{Discussion and Limitations}
Taken together, the results suggest a specific view of skills in multimodal agents. A useful skill should not remain an ephemeral prompt that is rewritten from scratch in each episode. Instead, it should be externalized as a procedural object that can be checked, repaired, and reused. This perspective helps explain why \method\ performs well across both VWA and MiniWoB. The method does not simply ask the model to ``try again,'' but converts failure into a structured repair problem over an explicit artifact.

At the same time, our current scope is intentionally narrow. We study repair for \rev{single} multimodal web skill artifacts rather than a full lifelong skill library. The artifact schema is tailored to browser interaction and does not yet cover the broader action space of desktop environments such as OSWorld or Agent S \cite{xie2024osworld,agashe2024agents}. Our transfer experiments test cross-model reuse within the same benchmark family, which is a meaningful but still limited notion of portability. In addition, deterministic verification is strongest when success can be expressed through programmatic checks over URL, DOM state, or form values  visually ambiguous or highly dynamic web pages remain harder to verify reliably. Finally, the released bundles are not uniformly instrumented for deeper analyses such as stability curves, patch-size distributions, or error-code breakdowns. For this reason, we center the paper on evidence that is consistently available across all settings, namely success, interaction cost, and cross-model reuse. We view richer longitudinal skill-bank analysis as a natural next step rather than a prerequisite for the present claim.

\section{Conclusion}
\rev{In this paper, we present } \method{}\rev{, a framework that turns a draft textual skill into a contracted executable artifact and improves it through deterministic verification, step-level fault localization, and minimal patch repair. Across both VisualWebArena and MiniWoB, } \method{} \rev{consistently outperforms self-generated skills, while the repaired artifacts remain reusable across GLM-4.6V and Qwen3.5-Plus. Taken together, these results suggest that skills in web agents should be treated as external procedural objects that can be audited, repaired, accumulated, and shared across models. We hope this perspective helps move multimodal web agents from one-shot prompting toward maintainable skill libraries.}

\bibliographystyle{ACM-Reference-Format}
\bibliography{refs}

\end{document}